\documentclass[letter,longauth,traditabstract]{aa} 

\usepackage{graphicx}
\usepackage{txfonts}
\usepackage{subfigure}
\begin{document}
\title{The dust morphology of the elliptical Galaxy M86 with SPIRE\thanks{{\it Herschel} is an ESA space observatory with science instruments provided by European-led Principal Investigator consortia and with important participation from NASA.}}


\author{
H.L. Gomez
\inst{1}
\and M. Baes \inst{2} 
\and L. Cortese \inst{1} 
\and M.W.L. Smith \inst{1} 
\and A. Boselli \inst{3}
\and L. Ciesla \inst{3}
\and G.J. Bendo \inst{4}
\and M. Pohlen \inst{1}  
\and S. di Serego Alighieri \inst{5} 
\and R. Auld \inst{1}
\and M.J. Barlow \inst{6} 
\and J.J. Bock\inst{7} 
\and M. Bradford \inst{7} 
\and V. Buat \inst{3} 
\and N. Castro-Rodriguez \inst{8} 
\and P. Chanial \inst{9}
\and S. Charlot \inst{10}
\and D.L. Clements \inst{4}
\and A. Cooray \inst{11}
\and D. Cormier \inst{9}
\and J.I. Davies \inst{1}
\and E. Dwek \inst{12} 
\and S. Eales \inst{1}
\and D. Elbaz \inst{9} 
\and M. Galametz \inst{9} 
\and F. Galliano \inst{9} 
\and W.K. Gear \inst{1} 
\and J. Glenn \inst{13} 
\and M. Griffin \inst{1} 
\and S. Hony \inst{9} 
\and K.G. Isaak \inst{14}
\and L.R. Levenson \inst{7} 
\and N. Lu \inst{7} 
\and S. Madden \inst{9} 
\and B. O'Halloran \inst{3}
\and K. Okumura \inst{9} 
\and S. Oliver \inst{15}
\and M.J. Page\inst{16}
\and P. Panuzzo \inst{9} 
\and A. Papageorgiou \inst{1} 
\and T.J. Parkin \inst{17}
\and I. Perez-Fournon \inst{8} 
\and N. Rangwala \inst{13} 
\and E.E. Rigby \inst{18} 
\and H. Roussel \inst{10} 
\and A. Rykala \inst{1} 
\and N. Sacchi \inst{19} 
\and M. Sauvage\inst{9}
\and M.R.P. Schirm \inst{17} 
\and B. Schulz\inst{20} 
\and L. Spinoglio \inst{19} 
\and S. Srinivasan \inst{10} 
\and J.A. Stevens \inst{21}
\and M. Symeonidis \inst{16} 
\and M. Trichas \inst{4} 
\and M. Vaccari \inst{22} 
\and L. Vigroux \inst{10} 
\and C.D. Wilson \inst{17} 
\and H. Wozniak \inst{23} 
\and G.S. Wright \inst{24} 
\and W.W. Zeilinger \inst{25}}

\institute{
School of Physics \& Astronomy, Cardiff University,
  The Parade, Cardiff, CF24 3AA, UK, \email{haley.gomez@astro.cf.ac.uk}
\and 
Sterrenkundig Observatorium, Universiteit Gent, Krijgslaan 281 S9, 
B-9000 Gent, Belgium
\and
Astrophysics Group, Imperial College, Blackett Laboratory, Prince 
Consort Road, London SW7 2AZ, UK
\and
Laboratoire d'Astrophysique de Marseille, UMR6110 CNRS, 38 rue F. 
Joliot-Curie, F-13388 Marseille France
\and 
INAF-Osservatorio Astrofisico di Arcetri, Largo E. Fermi 5, 
50125, Firenze, Italy
\and
Dept. of Physics and Astronomy, University College London, 
Gower Street, London WC1E 6BT, UK
\and
Jet Propulsion Laboratory, Pasadena, CA 91109, California Institute of Technology, Pasadena, 
CA 91125, USA
\and
Instituto de Astrof{\'\i}sica de Canarias (IAC) and Departamento de
Astrof{\'\i}sica, Universidad de La Laguna (ULL), La Laguna, Tenerife, Spain
\and
CEA, Laboratoire AIM, Irfu/SAp, Orme des Merisiers, F-91191, Gif-sur-Yvette, France
\and
Institut d'Astrophysique de Paris, UMR7095 CNRS, 98 bis Boulevard Arago, F-75014 Paris, France
\and
Dept. of Physics \& Astronomy, University of California, Irvine, CA 92697, USA
\and
Observational  Cosmology Lab, Code 665, NASA Goddard Space Flight  
Center Greenbelt, MD 20771, USA
\and
Department of Astrophysical and Planetary Sciences, CASA CB-389, 
University of Colorado, Boulder, CO 80309, US
\and
ESA Astrophysics Missions Division, ESTEC, PO Box 299, 2200 AG Noordwijk, The Netherlands
\and
Astronomy Centre, Department of Physics and Astronomy, University of 
Sussex, UK
\and
Mullard Space Science Laboratory, University College London, 
Holmbury St Mary, Dorking, Surrey RH5 6NT, UK
\and
Dept. of Physics \& Astronomy, McMaster University, Hamilton, 
Ontario, L8S 4M1, Canada
\and
School of Physics \& Astronomy, University of Nottingham, University 
Park, Nottingham NG7 2RD, UK
\and
Istituto di Fisica dello Spazio Interplanetario, INAF, Via del Fosso 
del Cavaliere 100, I-00133 Roma, Italy
\and
Infrared Processing and Analysis Center, California Institute of 
Technology, 770 South Wilson Av, Pasadena, CA 91125, USA
\and
Centre for Astrophysics Research, Science and Technology Research 
Centre, University of Hertfordshire, Herts AL10 9AB, UK
\and
University of Padova, Department of Astronomy, Vicolo Osservatorio 
3, I-35122 Padova, Italy
\and
Observatoire Astronomique de Strasbourg, UMR 7550 Universit\'{e} de 
Strasbourg - CNRS, 11, rue de l'Universit\'{e}, F-67000 Strasbourg
\and
UK Astronomy Technology Center, Royal Observatory Edinburgh, Edinburgh, EH9 3HJ, UK
\and
Institut f\"{u}r Astronomie, Universität Wien, Türkenschanzstr. 17, 
A-1180 Wien, Austria}
   \date{Submitted to A\&A Herschel Special Issue}

 
  \abstract
  { We present {\it Herschel}-SPIRE observations at 250-500\,$\mu$m of the
    giant elliptical galaxy M86 and examine the distribution of the
    resolved cold dust emission and its relation with other galactic
    tracers.  The SPIRE images reveal three dust components: emission
    from the central region; a dust lane extending north-south; and a
    bright emission feature $\rm 10\,kpc$ to the south-east. We
    estimate that $\rm \sim 10^6\,M_\odot$ of dust is spatially
    coincident with atomic and ionized hydrogen, originating from
    stripped material from the nearby spiral NGC~4438 due to recent
    tidal interactions with M86. The gas-to-dust ratio of the
    cold gas component ranges from $\sim 20-80$.  We
    discuss the different heating mechanisms for the dust features.  }


\titlerunning{{\it Herschel}-SPIRE observations of M86}

\authorrunning{H.L.Gomez et al.}

\maketitle

\section{Introduction}

Studies of cold dust in elliptical galaxies has been
limited to date by the lack of high-resolution, long wavelength
spectral coverage. In particular, the origin of far-infrared (FIR)
emission in these systems is still a controversial issue, with
evidence of dusty disks favouring a stellar origin (e.g. Knapp et al.\
1989) and other systems originating from mergers with dust-rich
galaxies (Leeuw et al.\ 2008).  The unprecedented resolution and
sensitivity of the recently launched {\it Herschel} Space Observatory
(Pilbratt et al.\ 2010) allows us to address long-standing issues such
as the origin and quantity of dust in ellipticals.

One of the most well-known IR-bright ellipticals is the giant Virgo
Cluster member, M86, at a distance of $\rm 17 \,Mpc$ (Mei et
al. 2007).  Two dust features were detected with {\it IRAS}, one
coincident with the galaxy and another to the north-west, thought to
be coincident with an X-ray plume of gas (Knapp et al.\ 1989; White et
al.\ 1991) and originally attributed to dust stripped from M86 due to
its motion through the cluster.  Higher resolution data from {\it ISO}
revealed two dust peaks within M86 suggesting a massive dust complex
(Stickel et al.\ 2003, hereafter S03).  They proposed a tidal origin,
also supported by absorption features attributed to dust stripped from
the nearby dwarf galaxy, VCC~882 (Elmegreen et al.\ 2000).  The
discovery of atomic gas offset from the centre of M86 and decoupled
from its stellar disk supports the tidal scenario (Li \& van Gorkom
2001).  More recently, Kenney et al.\ (2008, hereafter K08) detected
strong $\rm H{\alpha}$ features extending from the nearby spiral
NGC~4438 at $23\arcmin$ away, to within $1\arcmin$ of the nucleus of
M86.  The distribution and velocity of the ionized gas provides clear
evidence for tidal interaction between these two giants (K08, Fig.~1
Cortese et al.\ 2010).  In this scenario, we are observing debris from
the spiral left in the wake of the collision with M86 with $\sim
10^9\,\rm M_\odot$ of cold gas removed from NGC~4438 (K08).  The
stripped material is then heated by the hot interstellar medium (ISM)
of M86 or shock fronts from the interaction.  Although tidal
  stripping from NGC~4438 is supported by the atomic and ionized gas
  distribution, the origin of the dust responsible for the FIR
  emission and its heating mechanism is unclear.  Here we present
submm observations of M86 with {\it Herschel}-SPIRE (Griffin et
al. 2010). Companion observations of NGC~4438, are presented in
Cortese et al.\ (2010).

\section{Observations and data reduction}
\label{sec:obs}

\begin{figure*}
   \centering 
\includegraphics[trim=0mm 0mm 0mm
   0mm,clip=true,width=18cm]{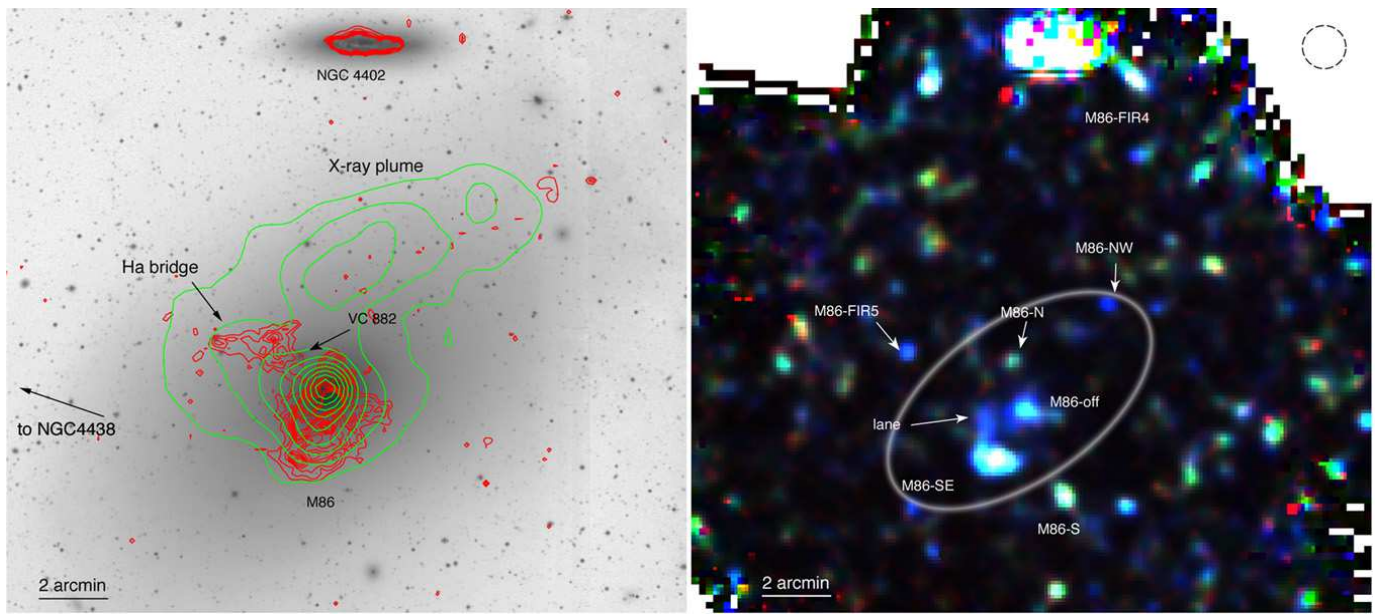}
   \caption{{\it Left:}  $R$-band image of M86 region with $\rm
     H\alpha$ (red) and X-ray (green) contours.  {\it Right:}
     Three-colour image at 250, 350 and 500\,$\mu$m, smoothed to
     500\,$\mu$m beam (as indicated by dashed circle in upper right of
     the figure).  The ellipse indicates the optical halo of M86
     ($8.9\arcmin \times 5.6\arcmin$). } 
         \label{fig:3col}
   \end{figure*}

   M86 was observed with SPIRE at 250, 350 and 500\,$\mu$m during {\it
     Herschel's} Science Demonstration Phase as part of the {\it
     Herschel} Reference Survey (Boselli et al.\ 2010a).  Eight pairs
   of cross-linked observations were taken in scan-map mode with
   scanning rate $30\arcsec$/sec. The data were processed following
   the detailed description given in Pohlen et al.\ (2010) and Bendo
   et al.\ (2010a).  Calibration methods and accuracies are outlined
   in Swinyard et al.\ (2010).  The measured $1\,\sigma$ noise level
   is 5, 6 and 7\,mJy\,beam$^{-1}$ at 250, 350 and 500\,$\mu$m with
   beam size 18, 25 and 37$\arcsec$; the noise is dominated by
   background source confusion.

\section{Results}
\label{sec:results}
\begin{figure*}
   \centering
\includegraphics[trim=0mm 54mm 0mm 0mm,clip=true,width=\textwidth]{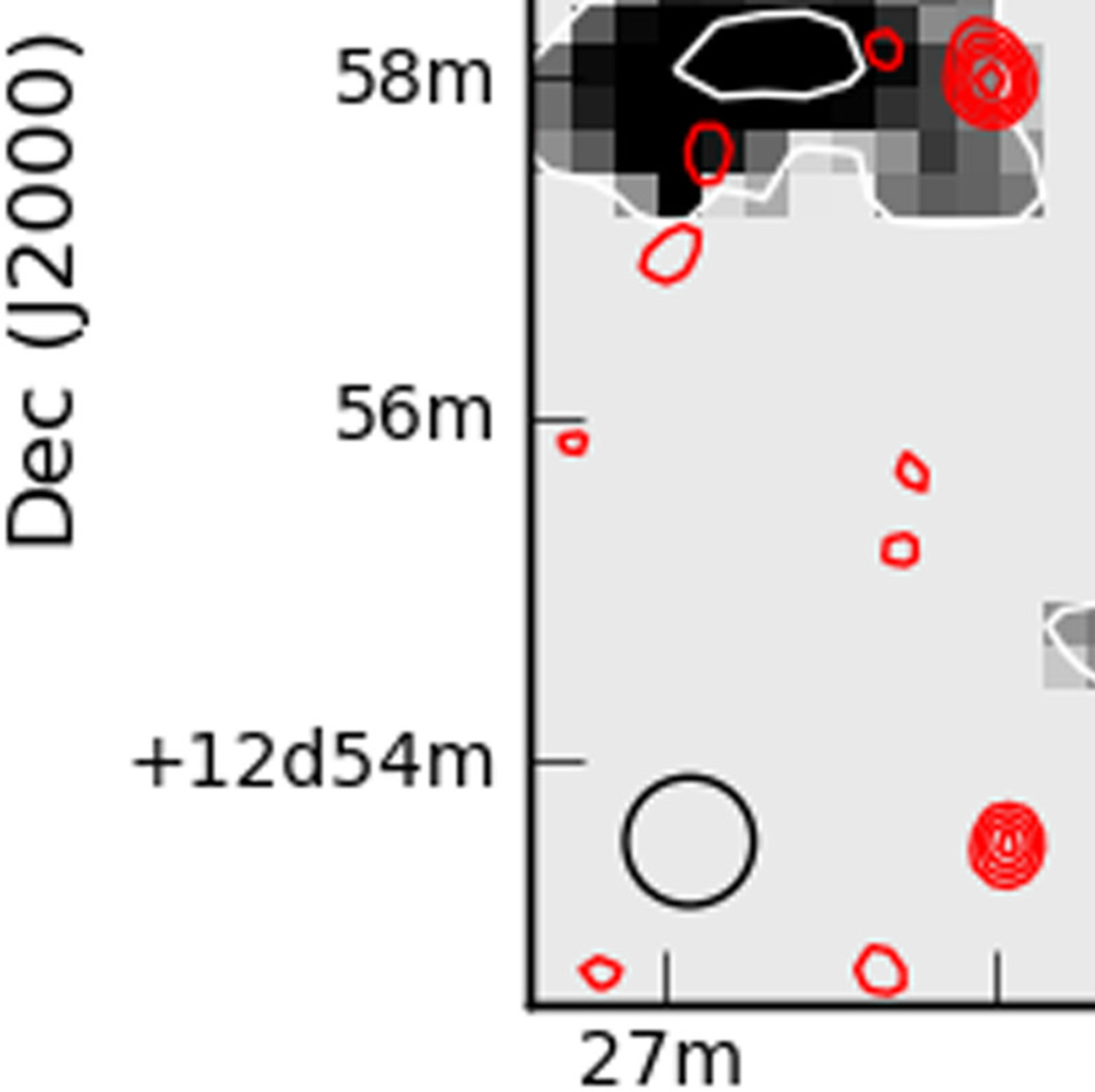}
   \caption{Multiwavelength comparison of M86.  {\it Top:} UV ({\it
       GALEX}) and optical (SDSS) with absorption features A \& B.
     {\it Middle:} {\it Spitzer} 24-160\,$\mu$m and SPIRE
     250-500\,$\mu$m. White ellipses indicate photometry apertures.
     {\it Bottom:} Gas emission shown in negative greyscale.  H{\textsc i}
     with white contours $\rm 10 + 25\,mJy\,beam^{-1}$ (courtesy 
     J. van Gorkom). $\rm H{\alpha}$ (courtesy J. Kenney) and X-ray
     ({\it ROSAT}).  250\,$\mu$m shown by red contours
     $9+6$\,mJy\,beam$^{-1}$.}
         \label{fig:grid}
   \end{figure*}

The three-colour SPIRE image is shown in Fig.~\ref{fig:3col} with dust
features labeled (following the terminology of S03) along with an
optical image of the same region with X-ray and $\rm H\alpha$
contours. In the SPIRE image, there are a number of unresolved
sources, and at least five extended features.  Careful comparison of
large scale FUV and IRAS maps of this region show that the cirrus
emission is extremely low ruling out a Galactic origin.  In
Fig.~\ref{fig:grid}, we focus on the central $13\arcmin \times
8\arcmin$ region and compare with structures seen at other
wavelengths.  Towards the south, M86-S is coincident with a number of
clustered 24\,$\mu$m sources.  In the north-west, the bright source
M86-NW (originally associated with M86's X-ray plume) has no optical
or UV counterpart and peaks at $\lambda <100\,\mu$m. The
bright feature M86-FIR5 is coincident with an optical galaxy (VPC~463)
and, like M86-FIR4, has a similar flux ratio to M86-NW.  Extending
north from M86, a faint ($2$-$3\sigma$) filament-like submm structure
appears to be coincident with blueshifted $\rm H\alpha$ emission
(Fig.~\ref{fig:grid}) attributed to ionized debris from the {\it
  incoming} trajectory of the NGC~4438 collision (Trinchieri \& di
Serego Alighieri 1991; Finoguenov et al.\ 2004; K08).  However, this
submm filament is also coincident with a number of distant
24\,$\mu$m sources.  We therefore suggest that this feature along with
M86-S, M86-NW, M86-FIR5 and M86-FIR4 originate from background sources
unrelated to M86.
  
   The submm morphology within the inner $2\arcmin$ of M86 is
   surprisingly complex and differs from the smooth distributions seen
   in the 24\,$\mu$m, UV and optical images.  We see four
     distinct features in the SPIRE maps labeled as M86-N, M86-off,
     M86-SE and `lane' (Fig.~\ref{fig:3col}).  M86-N is a bright
   unresolved source which is included in the photometry due to its
   coincidence with a $\rm H\alpha$ knot in this region\footnote{We
     note that M86-N has similar features to other point sources in
     the crowded field and could be a background object. It
     contributes 15\% of the flux measured in the aperture for
     M86-off.}.  The feature M86-SE ($7.7\times 2.3$\,kpc) is the
   brightest feature within the optical halo and is offset from the
   centre by $1.9\arcmin$ (10\,kpc). To the immediate north of this
   feature extends the bright `lane' structure.  These features are
   not seen in the {\it Spitzer} 24 or 70\,$\mu$m data (although the
   latter exhibits severe striping effects) but are detected with {\it
     ISO} at 90, 135 and 180\,$\mu$m and {\it Spitzer} 160\,$\mu$m
   (Fig.~\ref{fig:grid}).  Although there are four 24\,$\mu$m point
   sources coincident with the M86-SE structure, these are likely to
   be background sources since corresponding features are not seen in
   the same position in $\rm H{\alpha}$ nor are they coincident with
   the peaks in the submm emission.  These four features
   (Fig.~\ref{fig:grid}) are spatially coincident with peaks in the
   redshifted ionized gas (M86-SE: $-120\,\rm km\,s^{-1}$, M86-off:
   $-200-350\,\rm km\,s^{-1}$) extending from NGC~4438 (K08) and with
   atomic H{\sc i} (Li \& van Gorkom 2001).  The submm and H{\sc
     i} peak in M86-SE is coincident with an $\rm H\alpha$ `hole', but
   this is likely due to the lower resolution of the former.  M86-SE
   is also contained within an X-ray boundary tracing the X-ray halo
   (Finoguenov et al. 2004; Randall et al. 2008).  The strong
     spatial correlation between the cool atomic gas, cold dust,
     ionised gas and hot X-ray boundary shown in Fig.~\ref{fig:grid}
     seems to suggest that the dust originates from gas stripped from
     NGC~4438 and is immersed in the X-Ray halo of M86.

     At first glance, it appears that the `lane' feature and M86-SE
     are coincident with dusty filaments seen in absorption ({\it A}
     and {\it B} in Fig.~\ref{fig:grid}; Elmegreen et
     al. 2000).  Careful comparison with the SPIRE images show that
     their {\it B} feature is $1.3\arcmin$ south of M86-SE, and
     filament {\it A} is offset by $0.4\arcmin$.  Faint submm emission
     is seen at the southern tip of {\it A} and the atomic gas
     encompasses feature {\it B}.  We are not able to resolve {\it
       A} (at only $6\arcsec$ across in the optical), but
     surprisingly we do not detect significant emission associated
     with {\it B}.  This could be a result of the absorption features
     arising from foreground dust lanes with low column density, or
     the dust is simply too cold to be detected with {\it Herschel}.

\subsection{Dust mass and heating within the optical halo}
\label{sec:sed}
\begin{figure}
   \centering
   \includegraphics[trim=0mm 8mm 0mm 0mm,clip=true,angle=0,width=9.5cm]{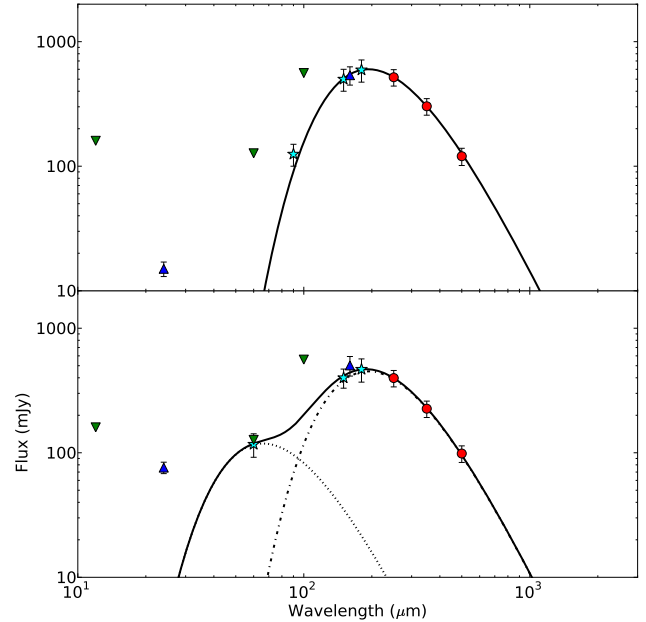}
   \caption{The SEDs for {\it top:} M86-SE and {\it bottom:} M86-off.
     Solid lines are modified blackbody functions with cold
     temperature $\rm 15\,K$ and $\rm 44\,K$.  {\it Spitzer} fluxes are blue
     triangles, {\it ISO} (stars) and SPIRE (circles). IRAS data
     (green triangles) are measured in a single aperture of radius
     $136\arcsec$.  }
         \label{fig:sed}
   \end{figure}

We performed aperture photometry on the IR and submm images from {\it
  ISO}, {\it Spitzer} and SPIRE.  The datasets were {\it wcs}-aligned,
and smoothed to the 160\,$\mu$m beam using the appropriate kernels
(Bendo et al.\ 2010b) and assuming gaussian beam profiles for SPIRE.
Fluxes were measured using two elliptical apertures
(Fig.~\ref{fig:grid}) with semi-major, semi-minor axes and position
angles of $1.4\arcmin \times 1.0\arcmin$, $127.5^\circ$ (encompassing
M86-SE) and $1.0\arcmin \times 1.6\arcmin$, $31.5^\circ$ (encompassing
M86-off and M86-N) respectively. We used an empirical 160\,$\mu$m PSF
to determine the aperture correction (Young, Bendo \& Lucero 2009).
The spectral energy distributions (SEDs) of M86-off and M86-SE are
shown in Fig.~\ref{fig:sed} (see also Boselli et al.\ 2010b).
Blackbody functions modified with a $\lambda^{-2}$ emissivity law are
plotted for comparison with temperatures $\rm 15\,K$ and $\rm 44\,K$.
Assuming a mass-absorption coefficient, $\kappa_{\rm 500} = 0.1\rm
\,kg^{-1}\,m^2$ (Draine 2003), we can estimate a rough dust mass from
the total 500\,$\mu$m flux, $S_{500}$ assuming $M_{\rm
  d}=S_{\nu}D^2/\kappa_{\nu}B(\nu, T)$ where $D$ is the distance. The
total dust mass within the inner $2\arcmin$ of M86 ranges from $\sim
\rm 2-5 \times 10^6 \rm \,M_{\odot}$ for temperatures, $T_{\rm d}
=15-20\,\rm K$.  Comparing the dust masses with the atomic gas mass,
we estimate a range of gas-to-dust ratios of $g/d \sim 20-80$
depending on the temperature. Non-detection of CO gas within M86 rules
out a significant molecular component with $< 10^{6.8}\,\rm M_{\odot}$
(Braine, Henkel \& Wiklind 1997). The ionized gas could contribute a
further $10^7\,\rm M_{\odot}$, but this is difficult to estimate
without knowing the gas densities (K08).  These ratios are higher than
expected in elliptical galaxies (e.g. Fich \& Hodge 1993) and similar
to the lower range of values estimated for the dust feature extending
out from NGC~4438 (Cortese et al.\ 2010).
   
It is interesting to ask what is responsible for heating the dust and
creating the submm emission seen here.  Possible mechanisms include
AGN, the interstellar radiation field (ISRF), tidal heating and/or the hot
X-ray halo.  Following Thomas et al.\ (2002), the luminosity $L_h$
required to heat a cloud of dust with grain size $a$, total mass
$M_{\rm d}$, temperature $T_{\rm d}$ and Planck-averaged absorption
coefficient $<Q(a,{ T_{\rm d}})>$, is given by Eq.~\ref{eq:thomas}:
\begin{equation}  
L_h = \left({4\pi a^2 \over{m_d}}\right)<Q(a,{T_{\rm d}})>T_{\rm d}^4 \sigma M_{\rm d}
\label{eq:thomas}
\end{equation}
For $a< 0.1\,\mu$m at $20$\,K, we require $L_h < 10^9\,\rm L_\odot$.
M86 is radio quiet and we can rule out significant heating from the
AGN.  Although the dust is not spatially coincident with the
optical/UV, the ISRF could still be responsible for heating the dust
with $L_h$ comparable to the $B$-Band luminosity (Mei et al.\
2007). To investigate this, we used a radiative transfer
implementation of the Monte Carlo code {\sc skirt} (Baes \& Dejonghe
2002), which models the absorption, scattering and thermal emission of
circumstellar discs and dusty galaxies.  M86 was represented by a
flattened Sersic model based on parameters from the literature (Caon,
Capaccioli \& D'Onofrio 1993; Graham \& Colless 1997; Gavazzi et al.\
2005) and we used the global SED from Boselli et al.\ (2010b) for the
intrinsic model. During the Monte Carlo simulation, the mean intensity
of the radiation field is calculated from UV-MIR wavelengths at every
position in the galaxy. From this mean intensity, the equilibrium dust
temperature, $T_{\rm eq}$ of each species of dust grains is calculated
using energy balance.  We predict that $T_{\rm eq}$ for silicate
(graphite) grains at a distance of $1.9\arcmin$ from the nucleus is
$13$ ($19$)\,K; even though the submm peaks do not correlate with
optical/UV emission, the ISRF of M86 is sufficient to produce the
submm emission detected here.

The thermal energy provided by M86's hot, X-ray emitting ISM could
also provide a significant heating contribution by collisionally
heating the dust, with grains of $a<0.1\,\mu$m reaching $T_{\rm eq} <
18$\,K (e.g. Dwek 1987).  The tidally-heated component as traced by
$\rm H\alpha$ also cannot be ruled out as it provides comparable
energy to M86's thermal reservoir over the last 100\,Myr ($\sim
10^9\,\rm L_\odot$, K08).  Thus, the ISRF, X-ray halo and tidal
heating could all be contributing to heating the dust and it is
difficult to separate these processes using the UV-submm SED only.
However, the high resolution images presented here suggest that submm
emission only originates from regions where we detect atomic {\it and}
ionised gas. Submm emission is not detected from regions with just
atomic H{\textsc i} and X-ray emission, suggesting that the X-ray halo
alone is not responsible for heating the dust. The spatial
distribution of these tracers therefore favours a scenario in which
the submm emission originates from dust mixed with stripped atomic
material and is heated by the tidal interaction.

In summary, we present submm images of M86 which reveal a complex dust
morphology with emission detected $\rm 10\,kpc$ from the centre, towards the
southeast.  The unprecedented resolution of SPIRE has revealed, for
the first time, a strong spatial correlation between the cold dust
($\rm \sim 10^6\,M_\odot$, $g/d$ of $20-80$) and the warm ionized gas
in M86.  This result strongly favours a scenario whereby the dust in
M86 originates from material stripped from the nearby spiral NGC~4438.
We investigate the different heating mechanisms responsible for the
dust emission detected by {\it Herschel}.  Intriguingly, although we cannot
rule out the stellar radiation field of M86, the strong correlation
between submm and H$\alpha$ emission suggests the cold dust is heated
by the same mechanisms responsible for ionizing the gas stripped from
NGC4438.  If so, tidal heating is likely to be responsible for the
dust emission.  Further modeling is required to provide a definite
answer on the origin of the submm features revealed by SPIRE in M86.

   \begin{acknowledgements} We thank Jeff Kenney \& Jacqueline van
     Gorkom for providing electronic versions of their data. The
     images were produced with APLpy, thanks to Edward Gomez \& Eli
     Bressart.  We thank the referee for their constructive
     comments. SPIRE has been developed by a consortium of institutes
     led by Cardiff University (UK) and including Univ. Lethbridge
     (Canada); NAOC (China); CEA, OAMP (France); IFSI, Univ. Padua
     (Italy); IAC (Spain); Stockholm Observatory (Sweden); Imperial
     College London, RAL, UCL-MSSL, UKATC, Univ. Sussex (UK); and
     Caltech/JPL, IPAC, Univ. Colorado (USA). This development has
     been supported by national funding agencies: CSA (Canada); NAOC
     (China); CEA, CNES, CNRS (France); ASI (Italy); MCINN (Spain);
     Stockholm Observatory (Sweden); STFC (UK); and NASA (USA).

\end{acknowledgements}


\begin{thebibliography}{}

\bibitem[]{} \bibitem[Baes 
\& Dejonghe(2002)]{2002MNRAS.335..441B} Baes M., \& Dejonghe H.\ 2002, \mnras, 335, 441


\bibitem[]{}Boselli A., Buat V., Ciesla L.,  et al., 2010b, A\&A, this volume

\bibitem[]{}Boselli, A., Eales, S., Cortese, L., et al., 2010a, PASP, 122, 261

\bibitem[]{} Braine J., Henkel C., Wiklind T., 1997, A\&A, 321, 765

\bibitem[]{} Bendo. G., Wilson C.D., Pohlen M., et al., 2010a, A\&A, this issue

\bibitem[]{}Bendo G., et al., 2010b, MNRAS, 402, 1409


\bibitem[]{} Caon, N., Capaccioli, M., D'Onofrio, M.\ 1993, \mnras, 265, 1013

\bibitem[2010]{cortese} Cortese L., Bendo G.J., Boselli A., et al.,
  A\&A, 2010, this volume



\bibitem[]{} Draine B.T., 2003, Ann. Rev. Astr. Ap., 41, 241

\bibitem[]{} Dwek E., 1987, ApJ, 332, 812

\bibitem[2000]{elmegreen} Elmegreen D.M., Elmegreen B.G., Chromey F.R., Fine M.S., 2000,ApJ,120,733

\bibitem[]{} Fich M., \& Hodge P., 1993, ApJ, 415, 75

\bibitem[2004]{finog} Finoguenov A., Pietsch W., Aschenbach B., Miniati F., 2004, A\&A, 415, 415


\bibitem[Gavazzi et  al.(2005)]{} Gavazzi, G., Donati, A., Cucciati, O., Sabatini, S., Boselli, A., Davies, J., \& Zibetti, S.\ 2005, \aap, 430, 411 

\bibitem[Graham \& Colless(1997)]{} Graham, A., \& Colless, M.\ 1997, \mnras, 287, 221  

\bibitem[]{}Griffin et al.\ 2010, A\&A, this volume

\bibitem[2008]{kenney} Kenney J.D.P., Tal T., Crowl H.H., Feldmeier J., Jacoby G.H., 2008, ApJ, 687, L69 {\bf [K08]}

\bibitem[]{} Knapp G.R., Guhathakurta P., Kim D-W., Jura M.A., 1989, ApJS, 70, 329


\bibitem[]{} Mei S., et al.\ 2007, ApJ, 655, 144


\bibitem[]{} Leeuw L.L., Davidson J., Dowell C.D., Matthews H.E., 2008, ApJ, 677, 249 


\bibitem[2001]{vangorkom} Li Y. \& van Gorkom J.H., in ASP
  Conf. Ser. 240, in {\it Gas and Galaxy Evolution}, ed. J.E. Hibbard, M. Rupen, and J.H. van Gorkom, 2001, 637

\bibitem[]{} Pilbratt et al.\ 2010, A\&A, this volume

\bibitem[]{} Pohlen M., Cortese L., Smith M.W.L., et al., 2010, A\&A, this volume

\bibitem[2008]{randall} Randall S., Nulsen P., Forman W.R., Jones C., Machacek M., Murray S.S., Maughan B., 2008, ApJ, 688, 208


\bibitem[2003]{stickel} Stickel M., Bregman J.N., Fabian A.C., White D.A., Elmegreen D.M., 2003, A\&A, 397, 503 {\bf [S03]}

\bibitem[]{} Swinyard et al.\ 2010, A\&A, this volume


\bibitem[]{} Thomas H.C., Dunne L., Clemens M.S., Alexander P., Eales S., Green D.A., James A., 2002, MNRAS, 331, 853

\bibitem[1991]{trinch} Trinchieri G., \& di Serego Alighieri S., 1991, AJ 101, 1647


\bibitem[]{} Young L.M., Bendo G.J., Lucero D.M., 2009, AJ, 137, 2

\bibitem[]{} White D.A., Fabian A.C., Forman W., Jones C., Stern C., 1991, ApJ, 375, 35


\end{thebibliography}
\end{document}